# Recording and Reproduction of Pattern Memory Trace in EEG by Direct Electrical Stimulation of Brain Cortex

Andrey G. Shapkin (1), Michael V. Taborov (2) and Yuriy G. Shapkin (3)

*(1) The East Siberian Scientific Centre of the Siberian Branch of Russian Academy of Medical Science, Neurosurgical Department, Irkutsk, Russian Federation, neuro@inbox.ru*
*(2) Institute of Solar-Terrestrial Physics of the Siberian Branch of Russian Academy of Science. Russian Federation, Irkutsk.*
*(3) State Medical University, Department of Pharmacology. Russian Federation, Irkutsk.*

**Abstract**

This study demonstrates the capability of external signal recording into memory and the reproduction of memory trace of this pattern in EEG by direct AC electrical stimulation of rat cerebral cortex. Additionally, we examine shifts of the DC potential level related to these phenomena. We show that in the course of memory trace reproduction, consecutive phases of engram activation and relaxation are registered and accompanied by corresponding negative and positive DC shifts. The observed electrophysiological changes may reflect consecutive activation and inhibition phases of neural ensembles participating in engram formation.

## Introduction

Despite prior success in revealing mechanisms of memory via discoveries in the field of synaptic and neuronal plasticity [1, 2, 3, 4, 5], the question of a functioning of apparatus responsible for memory trace recording and extraction remains unanswered. Progress in this field is often slow due to the limitations of existing biological models and certain methodologies for the extraction of specific engram correlates [6, 7, 8]. Consequently, ambiguity remains in the functional understanding of forming neural ensembles and the roles of emotion and motivation within the processes of memorisation and reproduction [9, 10, 11].

Memory trace formation is considered an integration of two neuronal ensembles activated by external signals [8, 12, 13]. One can assume that long-term activation of a particular bound neuron group by indifferent external stimulus and subsequent short-term presentation of a more significant, remembered pattern will be accompanied by the establishment of specific connections between the primarily activated neural network and other neuronal structures sensitive to this pattern. Repeated presentation of a reference signal and the resulting activation of primary neural ensemble elements will cause the subsequent activation of neural networks sensitive to this pattern. Therefore, we were able to artificially model this process by recording and directly visualising the signal that has been recorded by memory trace formation.

In this article, we will discuss the technique of administering external signals to be recorded into memory, the reproduction of memory traces of the respective pattern in EEG by direct electrical stimulation of rat cerebral cortex, and changes of the direct current (DC) potential level related to these phenomena.

## Material and methods

### Animals and surgery

The experiments were performed on 5 outbred male rats (200-220 g) that were housed in Plexiglas cages (48×24×20 cm) and provided food and water ad libitum. All animals were acquired and cared for in accordance with the guidelines published in the Guide for the Care and Use of Laboratory Animals (National Institutes of Health Publications No. 85-23, Revised 1985).

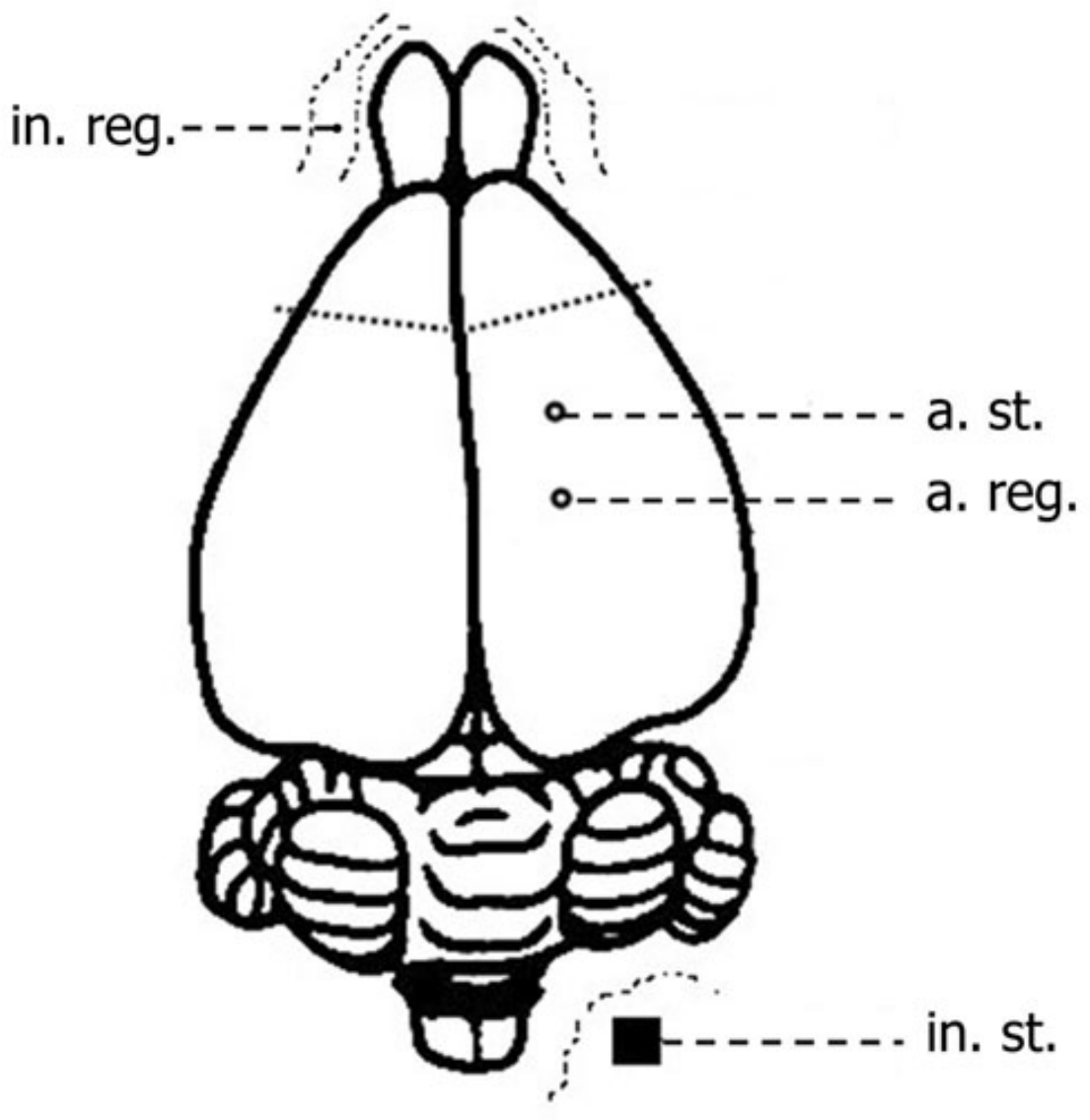

**Figure 1.** Schematic representation of electrode placement: a.st. –active stimulating electrode; in.st. –indifferent stimulating electrode (spinal muscles); a.reg. –active registering electrode; in.reg. –indifferent registering electrode (nasal bones).

Electrodes were implanted under sufficient anaesthesia, which consisted of an intraperitoneal injection of 50 mg/kg aethaminalum-natrium 20 minutes prior to the beginning of the operation. Non-polarising Ag/AgCl electrodes with a tip diameter of 0,25 mm were used and fixed with quick-setting acrylic plastic (Acrodent, "STOMA", Ukraine). The stimulating electrode was implanted in the epidural space at a point over the right parietal cortex, 3 mm to the right of the sagittal suture and 2 mm behind the coronal suture. The registering electrode was placed 2 mm behind the stimulating electrode. The indifferent registering electrode was fixed in the left skull nasal bones. The indifferent stimulating electrode (25 $mm^2$) was inserted on the spinal muscles (Fig. 1). The location of the electrodes was determined based on convenience of the implantation procedure, as prior single observations has not revealed any effect of stimulating electrode placement on experimental output. The placement of a registering electrode 3 mm from the stimulating electrode evoked an essential decrease in efficiency of a useful signal (pattern) extraction.

**Signals and Data collection**

The administration of a signal to be recorded into memory was carried out no sooner than 3 days after electrode implantation and involved repetitive, unipolar stimulation of the same point on the cortex surface with the reference (reading) signal and the basic (memorised) pattern. For the reference signal, we used a sinusoidal alternating current with a 7 Hz frequency for 30 seconds, and for the basic pattern, we selected an alternating current with a linear frequency increase from 0,5 to 10,5 Hz within 10 seconds (Fig. 2).

The control signal consisted of an alternating sinusoidal current with a frequency of 13 Hz for 30 seconds. The amplitudes of the reference and control signals were determined so that the registration point of the registered signal would not exceed 200 µV, and the ratio of the reference signal and basic pattern amplitudes would be 1:5. As previous single observations have shown, utilising signals with other amplitude-frequency characteristics (in the range from 0.5 to 15 Hz) does not influence experimental results, and they are primarily employed for the convenience of subsequent mathematical processing of the results.

Bioelectric activity was registered using the unipolar method with a DC amplifier with an input resistance of $10^8$ Ohm and a pass-band of 0-40 Hz. Bioelectric potentials were digitised with the frequency of 1024 Hz and entered into a computer system for further mathematical processing.

**Experimental stages**

The experimental scheme included the following 4 main components: 10-minute unipolar record of electrocorticogram, control stimulation of the cerebral cortex by the reference signal (7 Hz), and the learning and testing stages. The control stimulation procedure consisted of 10 applications of the randomly timed 7-Hz reference signal without subse-quent electrical stimulation using the basic pattern. The duration of electrical stimulation was 30 seconds, with a 60-second pause between stimulations. The learning procedure consisted of 25 consecutive, alternating applications of the reference signal and the basic pattern with randomly timed intervals between signals (no less than 1 and no more than 5 seconds). The testing procedure was conducted within 60 minutes of termination of the learning stage and consisted of random administration of reference (7 Hz) and control (13 Hz) signals. Each signal was applied at least 10 times during the testing procedure, with pauses between the signals lasting from 30 to 60 seconds.

### Data analysis and statistics

The analysis and averaging of infraslow brain electrical activity (DC potential) was conducted after the removal of the linear trend and application of an elliptic lowpass digital filter with a cutoff frequency of 0,15 Hz to single electrocorticogram fragments. Additionally, singled and averaged (for all animals) electrocorticogram fragments were visualised by means of time-frequency analysis using the Short-Time Fourier Transform method at each of the 4 experimental stages after preliminary band stop filtering of reference (7 Hz) or control (13 Hz) frequencies. The spectrogram was computed using the *specgram* function of the MATLAB Signal Processing Toolbox. To reduce noise, spectrograms were passed through 2-D adaptive noise-removal filtering with 15-by-15 neighbourhoods using the *wiener2* function from the MATLAB Image Processing Toolbox. Time parameters of the memory trace reproduction were calculated via the spectrograms' correlation of electrocorticogram and the basic pattern within the time-frequency range of 0,5-10,5 Hz and 0-10 seconds [14, 15]. Further averaging and comparison of statistically significant correlation coefficients was conducted using Fisher z-transformations. The detection threshold of the basic pattern was ±0.15, which was determined based on correlation coefficient limit fluctuations in the background EEG. We separately averaged the maximum values of correlation coefficients that exceeded modulo, the pattern detection threshold calculated from single electrocorticogram fragments and temporally corresponding amplitudes of infraslow electrical activity.

The visualisation of results and statistical and mathematical data processing were executed using MATLAB 7 and MS Excel 2003. A nonparametric U Wilcoxon-Mann-Whitney criterion was used for the evaluation of received output statistical significance. The differences are considered to be significant when $P<0.05$. Results are presented in the form of M ±m, where M is the mean and m is the error of the mean.

## Results

### Background EEG and control electrical stimulation stage

In the background electrocorticogram (50 non-artifactual fragments for 30 seconds), maximum infraslow electrical activity fluctuations did not exceed 200 μV and had an average value of 0.0±109.28 μV. Changes of the spectrogram's correlation coefficient of total background electrocorticogram and the basic pattern were in the limits of 0±0.11. During control electrical stimulation of cerebral cortex with the 7 Hz reference signal (50 fragments), the correlation coefficient fluctuated within the limits of 0±0.12 and did not significantly differ from similar changes in the background electrocorticogram. During electrical stimulation, DC potential tended to decrease by 50-100 μV from the initial level. On average, DC potential decreased by 28.94±78.29 μV during all stimulation period. These DC changes were not statistically significant in comparison to the initial level (before stimulation) and did not exceed physiological changes in the background EEG for the same period. The tendency towards a negative DC potential shift may have reflected a nonspecific response of the cerebral cortex to electric stimulation by the alternating current [16]. In spectrograms of electrocorticogram fragments recorded during this period as well as in the background electrocorticogram, no signal similar to the basic pattern (Fig. 2A, 3A) was visualised or determined using the correlation method.

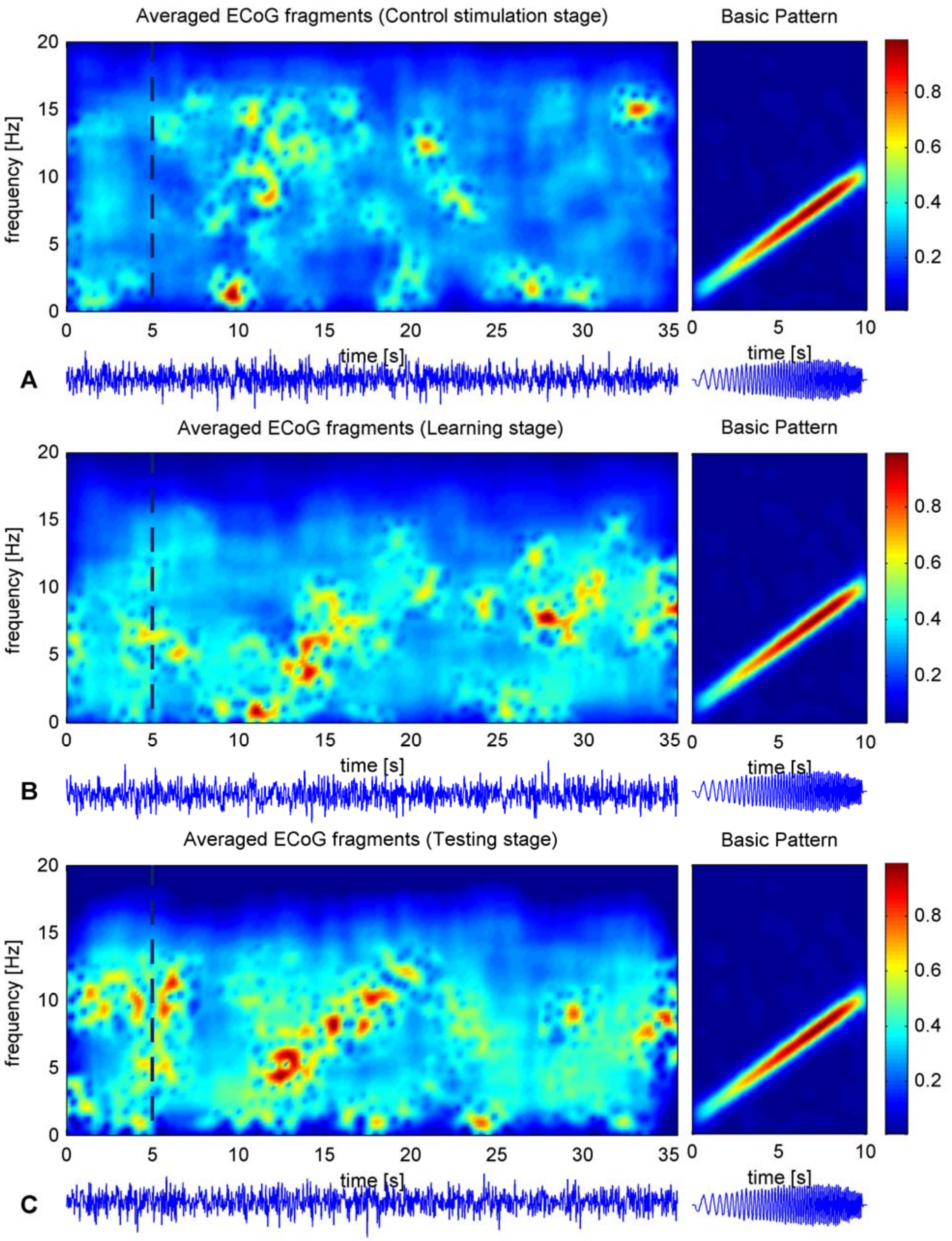

**Figure 2.** Spectrograms of averaged electrocorticogram fragments (after 7 Hz reference signal band stop filtering) during the stages of control stimulation (Fig. A, N=50), learning (Fig. B, N=125) and testing (Fig. C, N=50). The amplitude has been normalised, and the moment of reference signal application is designated by a dotted line. On all subfigures show the spectrogram of the basic pattern.

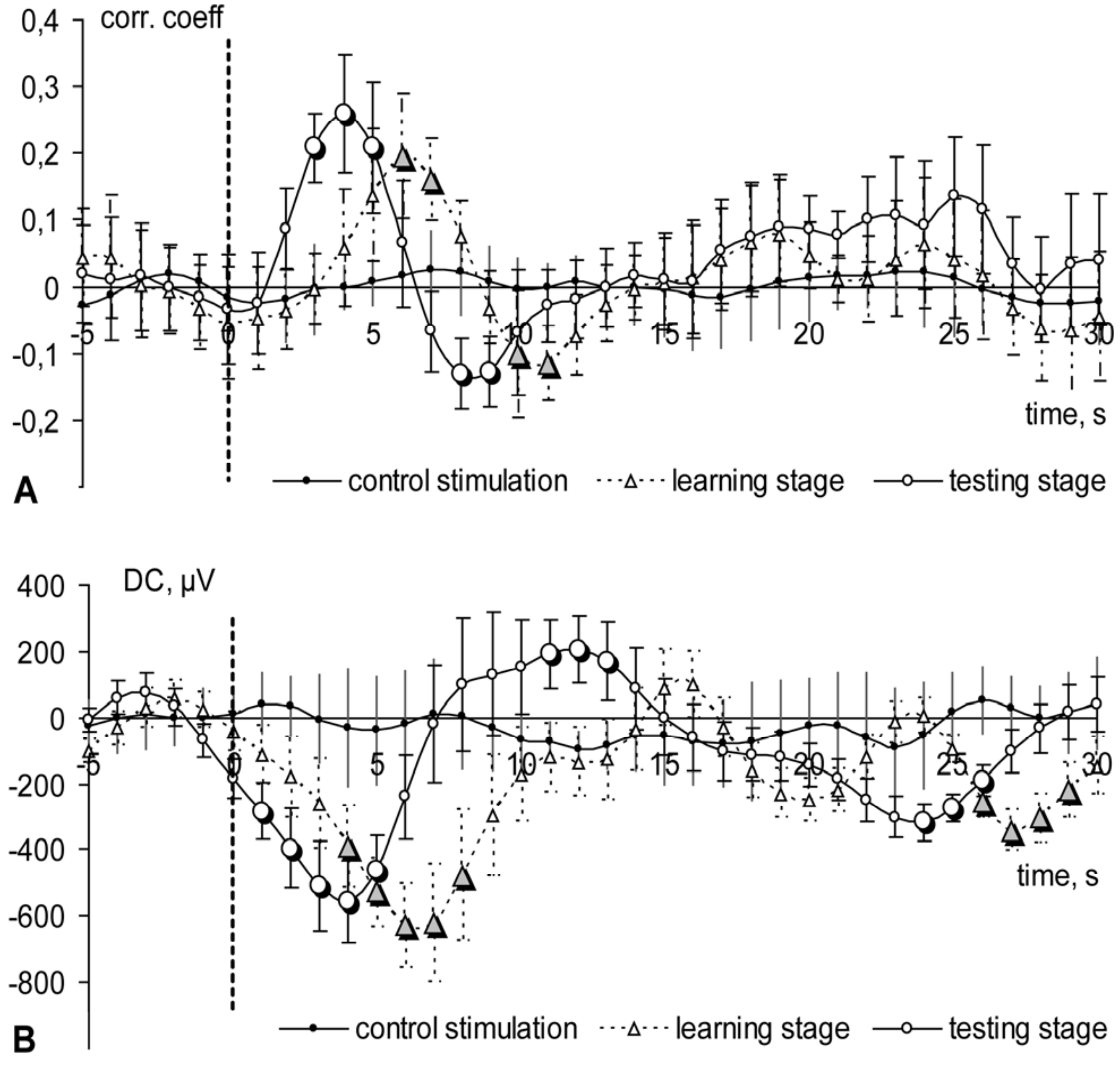

**Figure 3.** Correlation analysis results (Fig. A) and DC potential (Fig. B) during the stages of control stimulation (N=50), learning (N=125) and testing (N=50). Large markers show statistically significant changes (P<0.05) in comparison with the changes during control stimulation and the initial level (before stimulation).

**Learning stage**

As shown in Figure 2B, during the learning stage (125 fragments), a signal with time-frequency characteristics similar to those of the basic pattern was registered visually and via correlation analysis on 78 of 125 electrocorticogram fragments (62.4%) an average 6.2±2.3 seconds after the start of stimulation. This signal reproduction was referred to as an "engram activation phase." The spectrogram's correlation coefficient of averaged electrocorticograms (125 fragments) and the basic pattern during this period was 0.36. After averaging the correlation coefficient maximum values exceeding the detection threshold (78 fragments), the average maximum correlation coefficient of 0.26±0.11 was found (P <0.05 in comparison with the initial level) (Fig. 3A). The engram activation phase was accompanied by a negative DC shift of 508.39±187.67 µV at the moment when the correlation coefficient was at a maximum (78 fragments) (P<0.01 in comparison with the initial level) (Fig. 3B).

After primary signal generation, 67 of 125 (53.6%) spectrograms of electrocorticogram fragments showed consecutive suppression of the electrocorticogram amplitude with regards to time-frequency characteristics similar to those of the basic pattern an average of 10.8±3.5 seconds after reference signal application. These changes were accompanied by a positive DC shift of 135.47±82.98 µV from the initial level (67 ECoG fragments, P<0.01 in comparison with the engram activation period) (Fig. 3A). This phenomenon was referred to as an "engram relaxation phase". The spectrogram's correlation coefficient of averaged electrocorticogram fragments (125 fragments) and the basic pattern during this period was -0.26. After averaging the correlation coefficient minimum values exceeding the detection threshold (67 fragments), the average correlation coefficient of -0.21±0.04 was found (P<0.05 in comparison with the initial level, P<0.01 in comparison with the engram activation period) (Fig. 3B). In certain cases, several consecutive activation and relaxation phases were registered.

## Testing stage

### Reference signal (7 Hz) application

During test electrical stimulation with the 7 Hz reference signal 60 minutes after the learning procedure, the engram activation phase was registered 3.8±1.45 seconds after the beginning of electrical stimulation in 39 of 50 (78%) electrocorticogram fragments (Fig. 2C). This activation phase was accompanied by a negative DC shift of 417.07±237.87 µV (39 ECoG fragments, P<0.05 in comparison with the initial level) (Fig. 3B). The spectrogram's correlation coefficient of averaged electrocorticogram fragments (50 fragments) and the basic pattern during this period was 0.56. The average maximum correlation coefficient (39 fragments) was 0.28±0.12 (P<0.05 in comparison to the initial level) (Fig. 3A). The engram relaxation phase was registered 8.4±1.68 seconds after reference signal application in 31 of 50 (62%) electrocorticogram fragments and was accompanied by a positive DC shift of 219.04±104.77 µV (31 ECoG fragments, P<0.01 in comparison with the activation engram phase, P<0.05 in comparison with the control stimulation stage). The spectrogram's correlation coefficient of averaged electrocorticogram fragments (50 fragments) and the basic pattern during this period was -0.23, and the average correlation coefficient exceeding the detection threshold (31 fragments) was equal to -0.198±0.039 (P<0.05 in comparison with the initial level, P<0.01 in comparison with the engram activation period). We did not observe any statistically significant differences in correlation coefficients or DC potential changes during the engram activation or relaxation phases of the testing stage when compared to corresponding periods of the learning stage. In addition, despite a slight trend towards earlier engram activation and relaxation phases potentially associated with stably functioning newly formed neural networks, we did not note any statistically significant differences between the time of origin of these phases in the learning or testing periods.

### Control signal (13 Hz) application

Electrical stimulation during the testing period by the control signal (13 Hz, 50 fragments) did not exceed the basic pattern detection threshold, and correlation coefficient fluctuations averaged 0±0.11. Application of the 13 Hz control signal in testing stage, as well as control stimulation with the 7 Hz reference signal before the learning procedure, resulted in a tendency for the DC potential to decrease by an average of 44.83±116.08 µV from the initial level. Changes in the correlation coefficient and DC potential were not significant when compared to initial levels and did not differ significantly from corresponding changes in the background EEG.

## Discussion

Our results suggest that memory trace formation can be viewed as a process of integrating two neuronal ensembles activated by external signals [8, 12]. Activation of the neuronal system that is sensitive to the reference signal (7 Hz) as a result of long-term sensitisation by the alternating current [16] and the subsequent presentation of the basic pattern with complex time-frequency characteristics are associated with the creation of specific connections. These new communications exist between the primarily activated neural net-work and widely distributed neuronal structures sensitive to components of the basic pattern. Repeated application of the reference signal is associated with consecutive inclusion of these distributed neural networks, which are registered as a noise-like, frequency-selective reproduction of the electrocorticogram pattern (Fig. 2B, C). Negative and positive shifts of the DC potential during activation and relaxation phases may reflect consecutive depolarisation/repolarisation of the neuronal ensembles responsible for memory trace formation [17, 18, 19].

In our experience, DC potential changes during engram activation and/or relaxation can be a "cell-specific" analogue of an emotional component of memory trace formation and extraction processes. According to S.E. Murik [20, 21], the neurophysiologic foundation of emotions and motivations is tightly connected to polarising processes in sensory systems. Thus, deterioration of a neuron's functionality due to stimuli from the external and internal environments of an organism, reflected by membrane depolarisation and subsequent increases in metabolic and energy demands, is the manifestation of a negative emotional reaction. The reverse of this process is seen as an improvement of the functional condition of these neurons and a restoration of their membrane potential and is associated with a positive emotional reaction. Similarly, hyperpolarising processes accompanied by a positive DC shift are also connected to positive emotional reactions. Motivations are based on the desire to avoid an adverse functional state. According to the logic of this concept, in order to realise motivational behavior, the natural or artificial depolarisation of neuronal structures must promote the formation and extraction of memory traces. Presumably, this will result in a simplification of the engram activation phase and opposition to the relaxation phase [18, 22, 23, 24, 25]. The excessive depolarization under the influence of injurious factors such as ischemia, hypoxia, electric shock, KCl application and other [26, 27] must be associated with an inability to initiate the engram activation phase. This is due to the development of an adverse functional state, which manifests as depolarising inhibition of the neuronal activity [28]. These changes, caused by the physical or functional destruction of connections between components of neuronal structures of the reference signals and pattern, can explain memory consolidation or reconsolidation failures after similar pathological influences [29, 30, 31]. Hyperpolarising processes in neuronal ensembles, accompanied by positive DC shifts also due to hyperpolarising inhibition [28], must reversibly complicate memory trace reproduction (engram activation phase) and facilitate the relaxation phase [32].

In reality, external stimuli activate various, sometimes considerably spatially split, neuronal ensembles to form a system of interconnected, distributed neuronal structures that react exclusively to specific external or internal stimuli. As a result of these interactions, a complex space-time memory trace for an event is formed. Repeated full or partial activation of the preceding neuronal ensemble by an external stimulus causes consecutive activation of the connected neuronal structures and eventually subsequent relaxation.

## Conclusions

This study demonstrated signal recording into memory via direct brain cortex electrostimulation is possible with long-term preliminary activation of brain structures by a reference signal with subsequent presentation of a recorded pattern. The repeated application of this reference signal is accompanied by the reproduction of the signal in the electrocorticogram, with time-frequency characteristics similar to those of the pattern. In the course of memory trace reproduction, consecutive phases of engram activation and relaxation are registered accompanied by corresponding negative and positive DC shifts. Time parameters and correlation coefficients of engram activation and relaxation phases can be objective criterion for neural network functioning during recording and reproducing information within the cerebral cortex. The presented model can be used to directly study mechanisms of memory in the nervous systems of highly developed animals and for designing neurofeedback systems.